\documentclass[11pt,twoside]{article}

\usepackage{asp2006}
\usepackage{epsf}
\usepackage{lscape}

\begin{document}
\title{Yields from AGB Stars and their Impact on the Chemical 
Evolution of Dwarf Galaxies}   

\author{Simone Recchi}  

\affil{Institute of Astronomy, Vienna University, 
       Vienna, Austria 
       \\
       INAF -- Osservatorio Astronomico di Trieste,  
       Trieste, Italy }

\begin{abstract} 
By means of 2--D chemodynamical simulations, we study the evolution 
of dwarf galaxies with structural parameters similar to IZw18 and 
to tidal dwarf galaxies.  Different sets of yields from 
intermediate-mass stars are tested, in order to discover which one 
best reproduces the observed chemical compositions (in particular 
for nitrogen).  Different choices of yields from intermediate-mass 
stars lead to differences of up to 0.3--0.6 dex, depending on the
assumptions.  It is also shown that, given the dependence of the
cooling function on the metallicity, the dynamics of galaxies is
also significantly affected by the choice of nucleosynthetic yields.
\end{abstract}

\section{Introduction}

   From a nucleosynthetic point of view, asymptotic giant branch (AGB)
stars are characterized by the nuclear burning of hydrogen and helium
in thin shells on top of an electron-degenerate core of carbon and
oxygen.  Nuclear production from AGB stars is characterized by an
important array of elements just above H and He (see contributions of
Busso and Lattanzio in this volume), in particular C and N.  The 
importance of AGB stars on the global chemical evolution of galaxies 
is therefore evident (see also the contribution of Tosi in this volume).  
As an example, we show in Fig.~1 (left side) a comparison between 
the evolution of a chemodynamical model with (right panels) and 
without (left panels) the contribution of ejecta from AGB stars.  
The model reproduces the main features of the gas-rich dwarf galaxy 
IZw18 and assumes two bursts of star formation separated by 300
Myr of quiescence.  Only the evolution after the second burst is
shown.  The oxygen is almost unaffected (most of the $\alpha$-elements
are produced by massive stars).  Looking at the evolution of C/O and
N/O, the difference between the AGB and non-AGB models, and the
disagreement between the non-AGB model and the observations
(represented by shaded areas in Fig.~1), are striking.  Clearly, AGB
ejecta are a fundamental ingredient of the chemical evolution of
galaxies.

\begin{figure}[ht]
 \vspace{-0.4cm}
 \epsfxsize=7cm \epsfbox{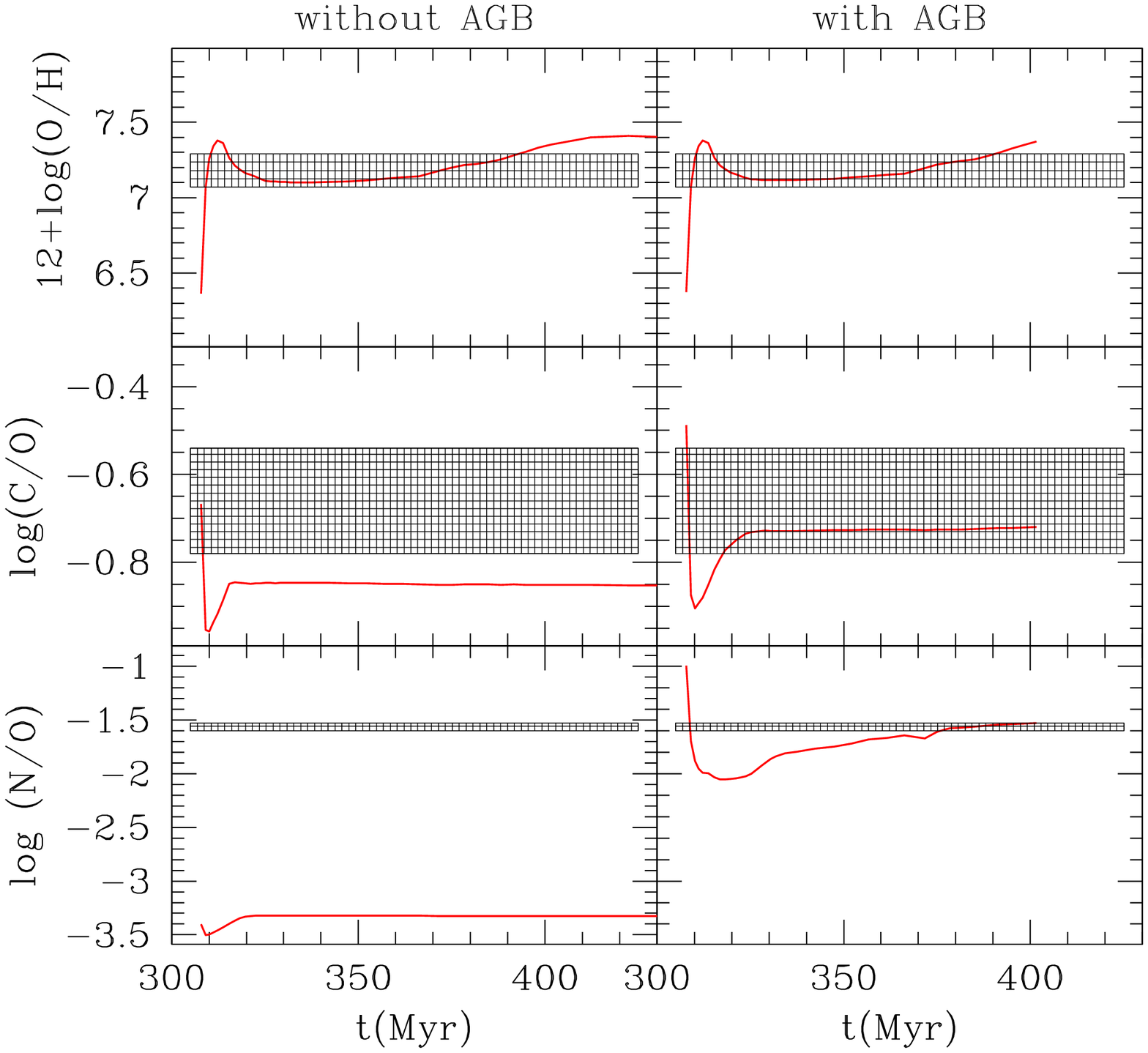} 
 \epsfxsize=7cm \epsfbox{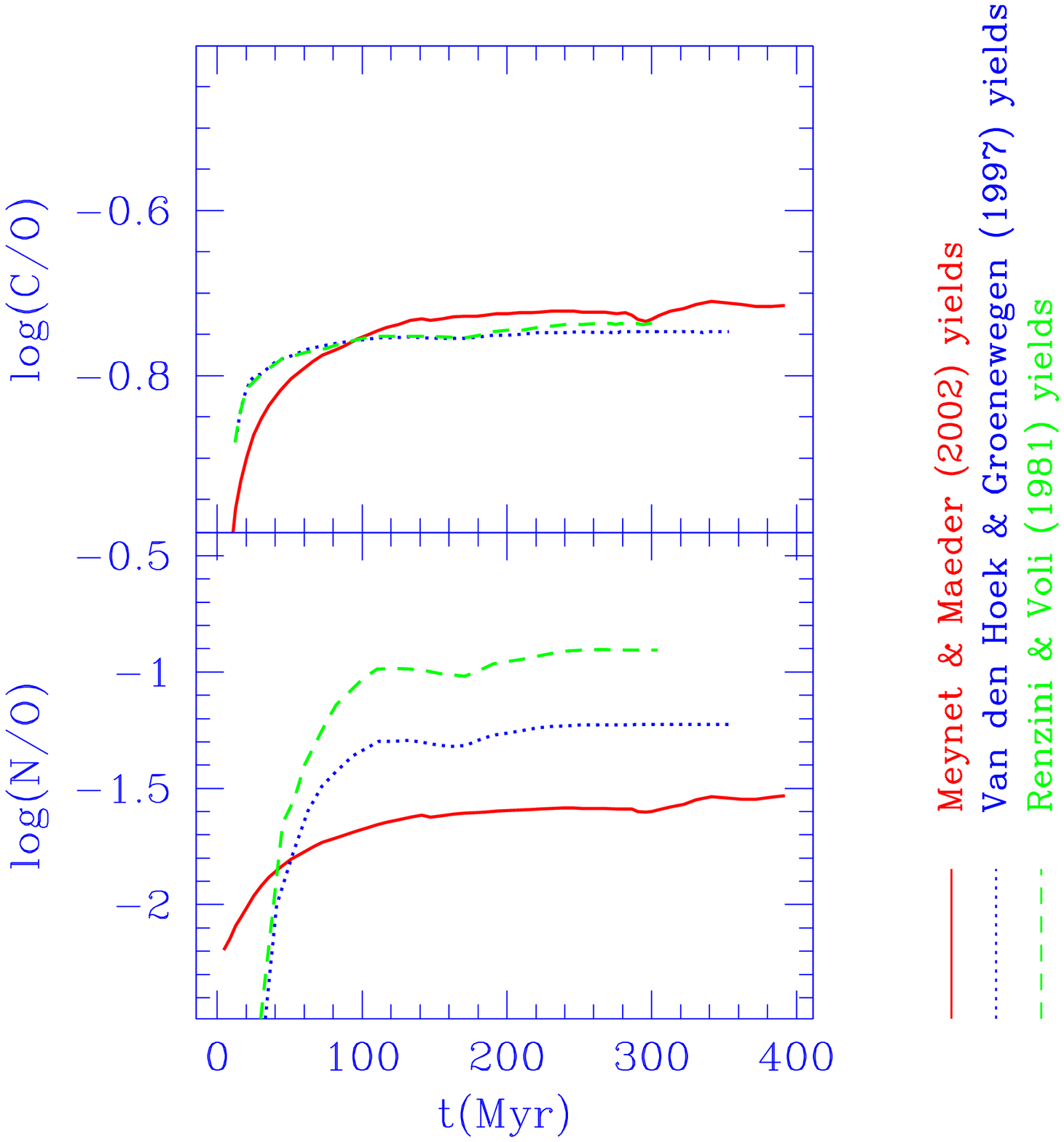} 
 \label{comp}
 \vspace{-1cm}
\caption{ ({\it Left side\/}): Evolution of 12 + log (O/H), log (C/O), 
  and log (N/O) for IZw18 models with ({\it right panels\/}) and 
  without ({\it left panels\/}) the contribution of AGB stars.  
  The shaded areas represent the observed values found in the
  literature for IZw18.  ({\it Right side\/}): log (C/O) and log
  (N/O) for IZw18 models assuming different nucleosynthetic 
  prescriptions for AGB stars (labeled to the right of the plot).}
\vspace*{-0.2cm}
\end{figure}

Gas-rich dwarf galaxies are characterized by peculiar chemical
compositions, hardly understandable in terms of closed-box models 
of chemical evolution.  In particular, the N/O ratios are puzzling.  
For 12 + log (O/H) larger than 7.8, the log (N/O) increases linearly 
with metallicity, although with large scatter.  This is consistent
with a secondary production of nitrogen.  At lower metallicities, 
all the galaxies seem to show a constant log (N/O) (of the order of
--1.55 to --1.60), with reduced scatter (Izotov \& Thuan 1999).  This 
is a typical behavior of elements produced in a primary way (namely,
starting from the C and O newly formed in the star).  Our aim in 
this contribution is to understand more about the time evolution of
particular abundance ratios.  We will explore specifically its
dependence on the adopted mode of star formation (SF) and on the
nucleosynthetic prescriptions.  We will show both models aimed at
reproducing IZw18 (\S~3.1) and models in which we simulate 
dark-matter-poor dwarf galaxies resulting from tidal interactions 
of large galaxies (\S~3.2).

\vspace*{-0.2cm}
\section{The Model}
\label{model}

In \S~3.1 we present models already published 
(Recchi et al.~2002, 2004), in which we simulate the main properties 
of the extremely metal-poor dwarf galaxy IZw18.  The adopted 
numerical code is described in the above-mentioned papers.  
Here we focus on the effect of nucleosynthetic prescriptions on the
chemical and dynamical evolution of the galaxies.  In \S~3.2 we
present preliminary results of new models in which
considerable improvements of the code have been implemented.  In
particular, the self-gravity has been taken into consideration, the
release of chemical elements from SNeII and SNeIa has been upgraded 
in order to take more properly into consideration primary and 
secondary contributions, SF now depends on the local thermodynamical 
characteristics of the gas, and the (metallicity-dependent) energy 
feedback from stellar winds is taken into consideration.

\vspace*{-0.2cm}
\section{Results}
\label{res}

\subsection{IZw18 Models}
\label{res_izw18}

We have considered (a) two instantaneous bursts of SF separated by
a quiescent period of 300 Myr, and (b) a SF history in which a long 
episode of SF of mild intensity is followed (after a short period 
without SF) by a more vigorous burst, lasting 5 Myr (``gasping" SF; 
see Recchi et al.~2004).  Three sets of yields from intermediate-mass 
stars (therefore three different prescriptions for the AGB ejecta) 
have been considered: Renzini \& Voli (1981 -- RV81); van den Hoek 
\& Groenewegen (1997 -- VG97), and Meynet \& Maeder (2002 -- MM02).  
In the first two cases, we combine these yields with the 
nucleosynthetic prescriptions for massive stars taken from Woosley 
\& Weaver (1995), whereas in the models adopting MM02 yields we 
adopt for consistency the whole set of masses calculated by those 
authors, ranging from 2 to 60 M$_\odot$.

The evolution of log (C/O) and log (N/O) for the IZw18 gasping models 
is shown in Fig.~1 (right side).  The dependence of log (C/O) on the
assumed nucleosynthetic prescriptions is rather small ($\sim$ 0.1 dex 
at most), whereas the difference in the nitrogen production is very 
remarkable.  In particular, MM02 already produce nitrogen in a primary
way in massive stars, so that the predicted N/O is larger than for
the other sets of yields in the first tens of Myr.  At later times,
when AGB stars start to dominate the chemical pollution of the galaxy,
the nitrogen predicted assuming MM02 yields is considerably lower than
the other models ($\sim$ 0.3 dex of difference assuming VG97 yields,
$\sim$ 0.6 dex less than what is predicted by the RV81 model).  It is
however worth recalling that the MM02 models do not take into
consideration later phases of the stellar evolution (in particular the
third dredge-up and the hot-bottom burning phase), N being mainly
produced in a primary way through rotational diffusion of C in the 
H--burning shell; therefore the predicted N/O should be considered as 
a lower limit.  Similar differences between various sets of yields are
also attained by the bursting models.  This is shown for instance in
Fig.~5 of Recchi et al.~(2002), where it can also be seen how little 
the different AGB yields affect the oxygen evolution.

The choice of MM02 yields gives better agreement between the model
results and the observed abundance ratios, in particular for log (N/O).  
Moreover, assuming a gasping SF regime, the log (N/O) attained for the 
MM02 model reaches $\sim$ --1.6 in about 120 Myr and then stays almost 
constant for the rest of the evolution of the object (Fig.~1, right 
side).  This behavior would easily explain the plateau at around this 
value observed in metal-poor dwarf galaxies.  A bursting SF would 
instead produce large variations of abundance ratios on short 
timescales which would easily lead to a large scatter in N/O.

\subsection{Models of Tidal Dwarf Galaxies}
\label{res_tdg}

Tidal Dwarf Galaxies (TDGs) evolve from self-gravitating structures
formed inside tidal tails, thrown out from interacting gas-rich
galaxies.  Theoretical arguments (Okazaki \& Taniguchi 2000) show 
that this mechanism of galaxy formation can account for most of the
present-day dwarf population.  However, from basic physical principles
and simulations it is clear that TDGs are characterized by a very
small dark matter (DM) content, so that a larger impact of the
feedback from the ongoing SF is expected.  Hence, the survival of TDGs
after many Gyrs is questioned.  We ran a set of chemodynamical models
aimed at simulating DM--poor dwarf galaxies (see Recchi et al.~2007)
in order to study under what conditions these objects can sustain the
energy released by dying stars without experiencing a blow-away.
As we have done in \S~3.1, we study here how these models are
affected by a different choice of AGB yields.

\begin{figure}[ht]
 \epsfxsize=7cm \epsfbox{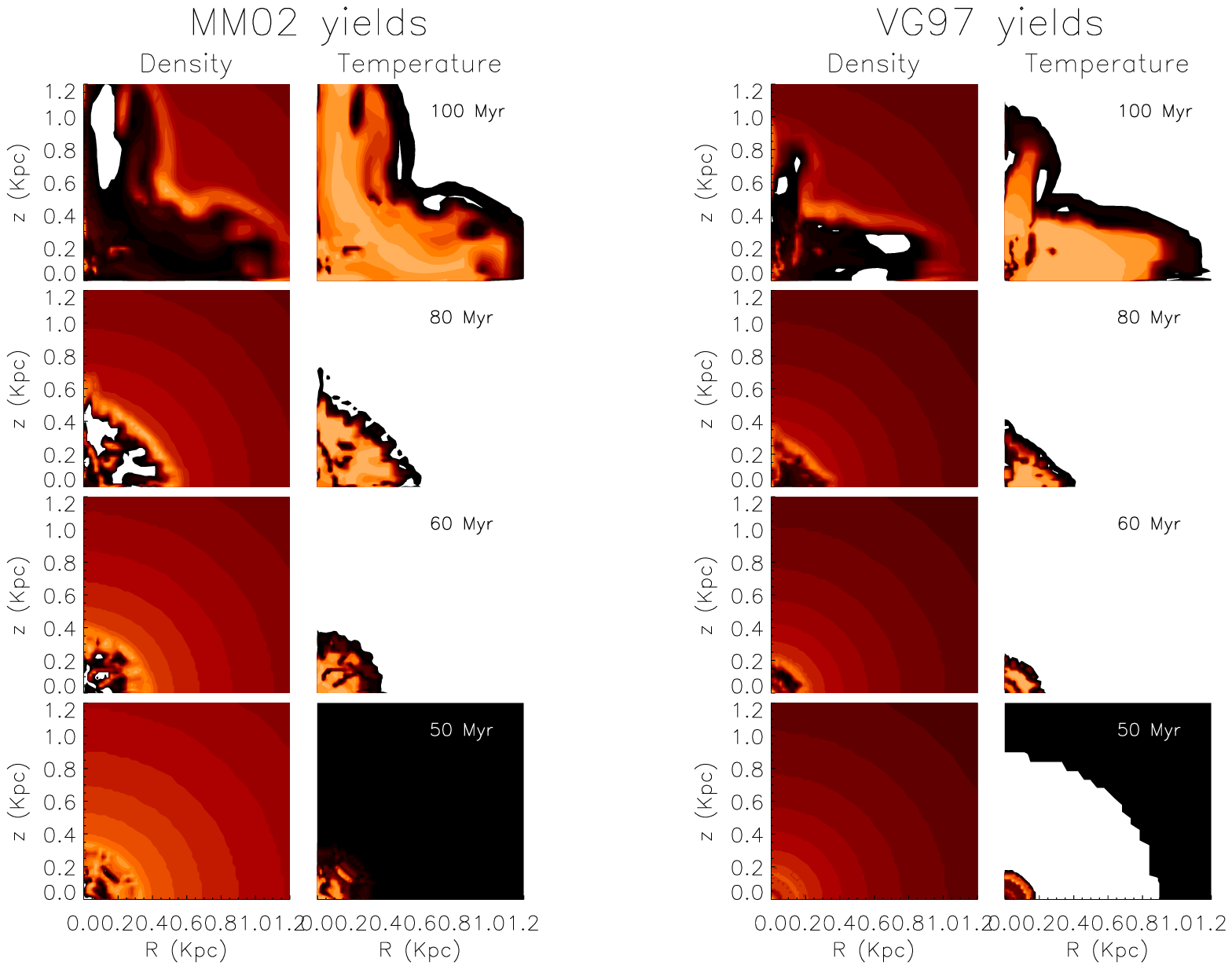} 
	\vspace{1cm}
 \epsfxsize=6cm \epsfbox{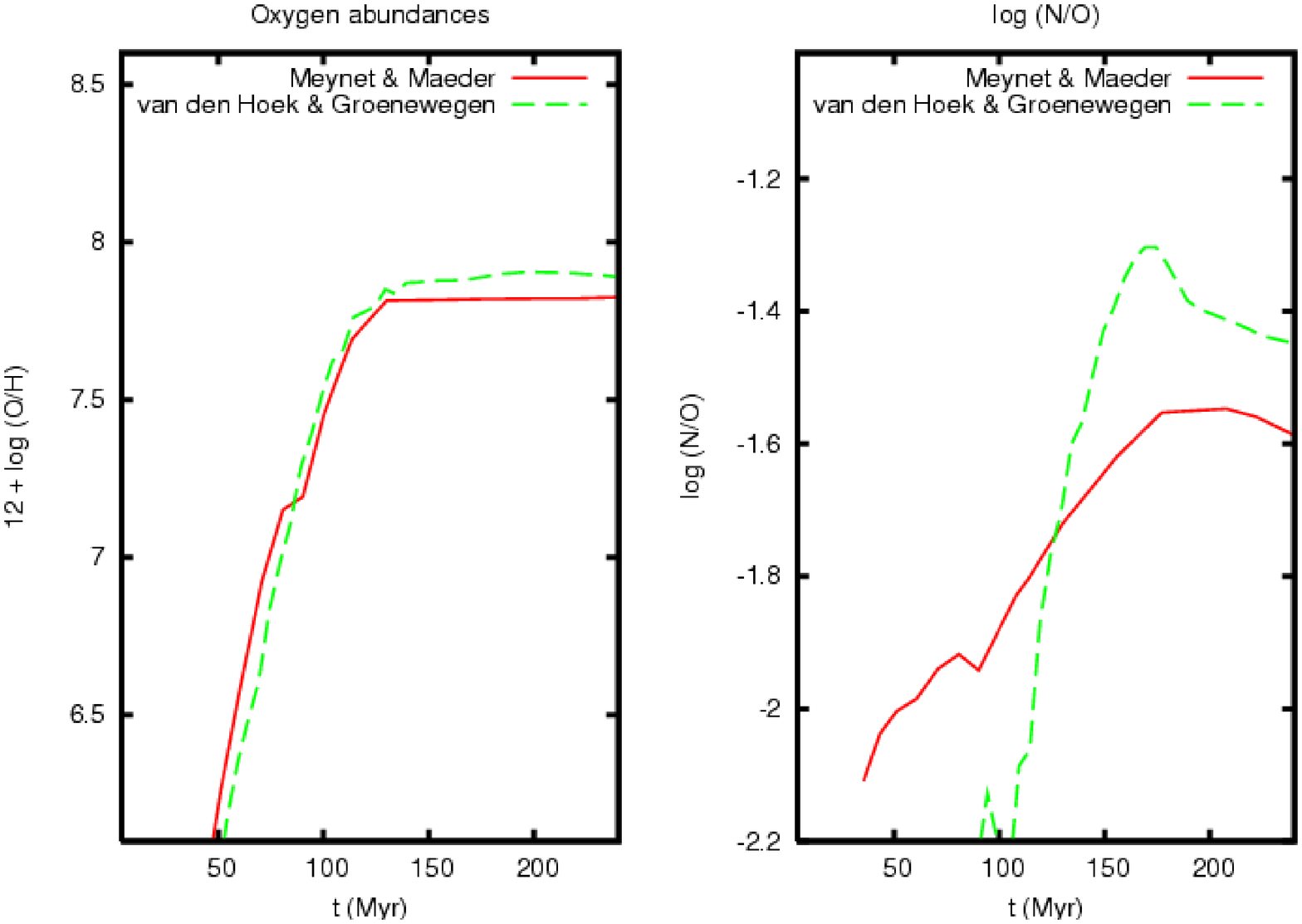}
	\vspace{-1cm} 
 \label{comp_tdg}
\caption{ ({\it Left side\/}): Density and temperature contours at 
           4 evolutionary times (labeled on each of the right panels) 
           for a DM--free model assuming MM02 ({\it left\/}) and VG97 
           ({\it right\/}) yields, respectively.  The (logarithmic) 
           density scale (in g cm$^{-3}$) ranges between --27 
           ({\it dark\/}) and --23 ({\it bright\/}).  The
           (logarithmic) temperature scale (in K) ranges between 3 
           and 7.  ({\it Right side\/}): Evolution of 12 + log (O/H) 
           ({\it left panel\/}) and log (N/O) ({\it right panel\/}) 
           as a function of time for the MM02 ({\it solid lines\/}) 
           and VG97 ({\it dashed lines\/}) models.}
\end{figure}

In Fig.~2 (left side) we show a comparison of the dynamical evolution
of two models differing only in the adopted nucleosynthetic
prescriptions: MM02 (left panels) and VG97 (right panels),
respectively.  These models have a SF efficiency $\varepsilon_{\rm
SF}$ = 0.2 and a temperature SF threshold $T_{\rm thr}$ = 10$^4$ K.
The superbubble evolution is faster in the MM02 model.  Indeed, MM02
produces on average more metals (in particular more
$\alpha$-elements), therefore leading to larger cooling rates.  On 
the one hand, it reduces the thermal energy content inside the 
superbubble, but on the other hand this increased cooling favors the 
process of star formation, leading to a more powerful feedback.  The
latter effect prevails, and a larger energy is available in model
MM02 to drive the expansion of the supershell.  In spite of the quite
evident differences, the gross features of the models are the same;
after $\sim$ 100 Myr a major galactic wind develops and most of the
gas is carried out of the galaxy.  On the right side of Fig.~2 we
compare the evolution of O/H (left panel) and N/O (right panel) as
a function of time for these two models.  The differences in the final
oxygen are very small, whereas looking at the evolution of N/O we
notice the same behavior that we pointed out in \S~3.1, namely 
log (N/O) is larger at the beginning for the MM02 model (due to a 
larger primary nitrogen production in massive stars), but when the 
nitrogen production from AGB stars becomes significant, the log (N/O) 
in the VG97 model overtakes the MM02 model, attaining in the end a 
value $\sim$ 0.2 dex larger.  These differences are therefore mostly 
due to the variations in the adopted sets of nitrogen yields rather 
than to the different dynamical behavior of the models.

The SF process in these models lasts $\sim$ 100 Myr.  Indeed, varying
parameters like $\varepsilon_{\rm SF}$ and $T_{\rm thr}$ it is
possible to produce stars at a milder rate, with less disruptive 
effects.  For instance, a model with $\varepsilon_{\rm SF}$ = 0.1 
or a model with a temperature threshold of $T_{\rm thr}$ = 10$^3$ K 
can produce stars at a rate of few tenths of a M$_\odot$ yr$^{-1}$ 
for at least 300 Myr.

\section{Conclusions}
\label{conc}

We have studied the dynamical and chemical evolution of dwarf galaxies
assuming different sets of yields from AGB stars.  We have focused
our study on two model galaxies: the first reproduces the main
structural properties of IZw18 while the second is a DM--poor model
aimed at studying the early evolution of tidal dwarf galaxies.  Our
main conclusions can be summarized as follows:

\begin{itemize}
	\vspace{-0.2cm}

\item Different sets of yields can affect the final predicted 
abundances and abundance ratios up to 0.5--0.6 dex, especially for 
nitrogen.
	\vspace{-0.2cm}

\item In the framework of gasping models of star formation, the 
abundance ratios are almost constant over large timescales, 
explaining the lack of a large spread in the observed log (N/O) 
of metal-poor dwarf galaxies.  
	\vspace{-0.2cm}

\item Due to the dependence of the cooling curve on the metallicity, 
models with different sets of yields also have different dynamical 
evolutions.
	\vspace{-0.2cm}

\item Models of DM--poor dwarf galaxies are not necessarily rapidly 
destroyed and can survive the feedback of the ongoing star formation 
for more than 300 Myr.
	\vspace{-0.2cm}

\end{itemize}

\acknowledgements 
The SOC is warmly acknowledged for accepting my contribution and for
putting together a very interesting conference.  Decisive
contributions to the work presented here have come from my 
collaboration with A.~D'Ercole, G.~Hensler, P.~Kroupa, F.~Matteucci, 
Ch.~Theis and M.~Tosi.  I also acknowledge financial support from the 
Deutsche Forschungsgemeinschaft (DFG) under grants HE 1487/28 and 
TH 511/8.

\question{Gustafsson} Could you comment on what would happen, in 
particular to your short mixing times, if you would allow a third 
spatial dimension?

\answer{Recchi} The third dimension would increase the turbulence 
in the ISM, therefore allowing for more efficient mixing of the newly 
produced metals.


\begin{thebibliography}{}

\bibitem[1999]{it99} Izotov, Y.\,I. \& Thuan, T.\,X. ~1999, ApJ, 511, 
         639

\bibitem[2002]{mm02} Meynet, G. \& Maeder, A. ~2002, A\&A, 390, 561 
         (MM02)

\bibitem[2000]{ot00} Okazaki, T. \& Taniguchi, Y. ~2000, ApJ, 543, 149

\bibitem[2002]{recc02} Recchi, S., Matteucci, F., D'Ercole, A., \& 
	Tosi, M. ~2002, A\&A, 384, 799

\bibitem[2004]{recc04} Recchi, S., Matteucci, F., D'Ercole, A., \& 
	Tosi, M. ~2004, A\&A, 426, 37

\bibitem[2007]{recc07} Recchi, S., Theis, Ch., Kroupa, P., \& 
	Hensler, G. ~2007, A\&A, 470, L5

\bibitem[1981]{rv81} Renzini, A. \& Voli, M. ~1981, A\&A, 94, 175 
        (RV81)

\bibitem[1997]{vg97} van den Hoek, L.\,B. \& Groenewegen, M.A.T. 
        ~1997, A\&AS, 123, 305 (VG97)

\bibitem[1995]{ww95} Woosley, S.\,E. \& Weaver, T.\,A. ~1995, ApJS, 
         101, 181

\end{thebibliography}
\end{document}